\documentstyle[amssymb,aps,prl,epsfig]{revtex}
\begin{document}
\draft
\twocolumn[\hsize\textwidth\columnwidth\hsize\csname %
@twocolumnfalse\endcsname

\title{Pseudogap and the Fermi surface in the $t$-$J$ model}
\author{A. Ram\v sak and P. Prelov\v sek}
\address{Faculty of Mathematics and
Physics, University of Ljubljana, 1000 Ljubljana, Slovenia }
\address{J. Stefan Institute, University of Ljubljana, 1001 Ljubljana,
Slovenia }
\date{June 30, 2001}    
\maketitle 
\begin{abstract}  
\widetext  
We calculate spectral functions within the $t$-$J$ model as relevant
to cuprates in the regime from low to optimum doping. On the basis of
equations of motion for projected operators an effective spin-fermion
coupling is derived. The self energy due to short-wavelength
transverse spin fluctuations is shown to lead to a modified
selfconsistent Born approximation, which can explain strong asymmetry
between hole and electron quasiparticles. The coupling to
long-wavelength longitudinal spin fluctuations governs the
low-frequency behavior and results in a pseudogap behavior, which at
low doping effectively truncates the Fermi surface.
\vskip 2cm
\end{abstract} 
\vskip 1cm]
\narrowtext
Here we concentrate on some of the experimental facts revealing the
nature of quasiparticles (QP) and pseudogap in underdoped
cuprates. Several quantities, in particular the uniform
susceptibility, the Hall constant, the specific heat, show the (large)
pseudogap scale $T^*$ \cite{imad}, consistent with the angle resolved
photoemission (ARPES) revealing a spectral function $A(\bf k,\omega)$
with a hump at $\sim 100$~meV observed in
Bi$_2$Sr$_2$CaCu$_2$O$_{2+\delta}$ (BSCCO) near the momentum ${\bf
k}=(\pi,0)$\cite{mars}. QP dispersing through the Fermi surface (FS)
are resolved by ARPES in BSCCO only in parts of the large FS, in
particular along the nodal $(0,0)$-$(\pi,\pi)$ direction, indicating
that the rest of the large FS are either fully or effectively gaped.
Some aspects of the pseudogap have been found in the phenomenological
spin-fermion models
\cite{pine}. The renormalization group studies of the Hubbard model 
\cite{zanc} also indicate on the breakdown of the standard Fermi liquid 
and on the truncation of the FS.  That such features also emerge from
prototype models of correlated electrons has been confirmed in
numerical studies of spectral functions in the Hubbard \cite{preu} and
in the $t$-$J$ model \cite{jpspec,jprev}, which both show the
appearance of the pseudogap at low doping.

Our aim is to capture these features within an analytical treatment of
a single band model. The most difficult aspect in the latter is
inherent strong coupling between mobile fermions and spin degrees, for
which it is hard to find even a proper phenomenological model.  In the
following we show that such an effective spin-fermion model can be
derived via equations of motion (EQM) and dividing the coupling into
short and long-wavelength spin fluctuations.

We study the planar $t$-$J$ model
\begin{equation}
H=-\sum_{i,j,s}t_{ij} \tilde{c}^\dagger_{js}\tilde{c}_{is} 
+J\sum_{\langle ij\rangle}({\bf S}_i\cdot {\bf S}_j-\frac{1}{4}
n_in_j) , \label{eq1}
\end{equation}
where we take into account possible longer range hopping, i.e.,
besides $t_{ij}=t$ for n.n. hopping also $t_{ij}=t'$ for n.n.n.
neighbors on a square lattice. The latter appears to be relevant in
the study as shown later. We evaluate the single-particle propagator
in this model explicitly taking into account that fermionic operators
are projected ones not allowing for double occupancy of sites, e.g.,
$\tilde{c}^\dagger_{is}= (1-n_{i,-s}) c^\dagger_{is}$.
We use EQM directly for projected operators \cite{prel} and
represent them in variables relevant for a paramagnetic metallic state
with $\langle {\bf S}_i \rangle =0$ and electron concentration $\langle n_i
\rangle = c_e=1-c_h$.
EQM for $\tilde c_{{\bf k} s}$ can be used to construct approximations
for the electron propagator $G({\bf k},\omega)$  \cite{prel,plak,pr01}, 
which can be represented as
\begin{equation}
  G({\bf k},\omega)= \frac{\alpha}{\omega+\mu -\zeta_{\bf k} - \Sigma({\bf
      k},\omega) }, \label{eq3}
\end{equation}
where the renormalization $\alpha= (1+c_h)/2$ is a consequence of a
projected basis, and $\zeta_{\bf k}= -4 \eta_1 t \gamma_{\bf k}
-4 \eta_2 t' \gamma'_{\bf k}$ is the 'free' propagation term
emerging from the EQM, with $\gamma_{\bf k}=(\cos
k_x+\cos k_y)/2$ and $\gamma_{\bf k}'=\cos k_x \cos k_y$. Here
$\eta_j = \alpha + \langle S_0^z S_j^z \rangle/\alpha$ is
determined by spin correlations.  The central quantity for further
consideration is the self energy $\Sigma({\bf k},\omega) =
\langle\!\langle C_{{\bf k}s};C^+_{{\bf k}s} \rangle\!\rangle_\omega^{irr} /
\alpha $, where $iC_{{\bf k}s}=[\tilde c_{{\bf k} s},H]-\zeta_{\bf k}
\tilde c_{{\bf k} s}$, and only the 'irreducible' part of the correlation
function should be taken into account in the evaluation of $\Sigma$.
In finding an approximation for $\Sigma$ we assume that we are dealing
with the paramagnet with pronounced AFM SRO with the dominant wave vector
${\bf Q} =(\pi,\pi)$ and the AFM correlation length $\xi>1 $ with
corresponding $\kappa =1/\xi$. EQM naturally indicate on
an effective coupling between fermions and spin degrees, however the role of
short-range and longer-range spin fluctuations is quite different.

In an undoped AFM system the spectral function of an added hole is
quite well described within the selfconsistent Born approximation
(SCBA) \cite{kane}, where the strong hole-magnon coupling induced by
the hopping $t$-term leads to a broad background representing the
incoherent hopping and a quite narrow QP dispersion determined by
$J$. Our EQM formalism naturally reproduces SCBA in an undoped system
and it is easy
to generalize the equations for finite doping $c_h>0$ where we have
also electron-like QP above the Fermi energy ($\omega>0$). In 2D the
AFM long-range order is broken at $T>0$ and $c_h>0$, still spin
fluctuations are magnon-like 
with a dispersion $\omega_{\bf q}$ for $q>\kappa$ and
$\tilde q>\kappa$ where $\tilde {\bf q}= {\bf q}-{\bf Q}$. The
paramagnon contribution to the self energy can be written as
\begin{eqnarray}
  &&\Sigma_{\rm pm}({\bf k},\omega)= \frac{16t^2}{N} \sum_{q,
\tilde q> \kappa}
  (u_{\bf q} \gamma_{{\bf k}-{\bf q}}+v_{\bf q} \gamma_{\bf k})^2
  \nonumber \\ &&[ G^-({\bf k}-{\bf q},\omega+\omega_{\bf q}) +
  G^+({\bf k}+{\bf q},\omega-\omega_{\bf q})], \label{eq5}
\end{eqnarray}
where $(u_{\bf q},v_{\bf q})=(1,-{\rm sign}(\gamma_{\bf q}))\sqrt{(2J
\pm\omega_{\bf q})/2\omega_{\bf q}}$ and $G^\pm$ refer to the Green's
functions corresponding to electron ($\omega>0$) and hole ($\omega<0$)
QP excitations, respectively.

We are dealing with a paramagnet without the AFM long-range order,
therefore it is essential to consider also the coupling to
longitudinal spin fluctuations which for $\tilde q<\kappa$
appear to be quite uncoupled, therefore we express the longitudinal
contribution as in Refs.~\cite{prel,plak,pr01},
\begin{eqnarray}
\Sigma_{\rm lf}({\bf k},\omega) &=&\frac{1}{\alpha N} \sum_{\bf q} 
\tilde m^2_{\bf k q}
\int \int \frac{d\omega_1 d\omega_2}{\pi} g(\omega_1,\omega_2) \nonumber\\
&&\frac{A^0({{\bf k}-{\bf q}},\omega_1) \chi''({\bf q},\omega_2)}
{\omega-\omega_1-\omega_2 }, \label{eq7}
\end{eqnarray}    
where $\chi({\bf q},\omega)$ is the dynamical spin susceptibility,
$A^0({\bf k},\omega)=-(\alpha/\pi){\rm Im}(\omega+\mu -\zeta_{\bf k} -
\Sigma_{\rm pm})^{-1}$,
$g(\omega_1,\omega_2)=[{\rm th}(\omega_1/2T)+{\rm cth}(\omega_2/2T)$]/2
and $\tilde m_{\bf kq}= 2J
\gamma_{\bf q}+ \frac{1}{2} (\epsilon^0_{{\bf k}-{\bf q}}+
\epsilon^0_{\bf k})$ with 
$\epsilon^0_{\bf k}=-4t\gamma_{\bf k}-4t'\gamma'_{\bf k}$\cite{pr01}.

In $\Sigma_{\rm lf}$ only the part corresponding to irreducible
diagrams should enter, so there are restrictions on proper
decoupling. We use in Eq.~(\ref{eq7}) the most appropriate and simplest
approximation to insert the unrenormalized $A^0({\bf k},\omega)$,
i.e., the spectral function without a self-consistent consideration of
$\Sigma_{\rm lf}$ but with $\Sigma_{\rm pm}$ fully taken into account.
Such an approximation has been introduced in the theory of a pseudogap
in CDW systems \cite{lee}, used also in related works analyzing the
role spin fluctuations \cite{kamp},\cite{chub}, and recently more
extensively examined in Ref.\cite{mill}.

So far we do not have a corresponding theory for the spin response at
$c_h>0$ and $T>0$, so $\chi({\bf k},\omega)$ is assumed as a
phenomenological input, bound by the sum rule
\begin{equation}
  {1\over N} \sum_{\bf q} \int_0^{\infty} {\rm
    cth}(\frac{\beta\omega}{2}) \chi''({\bf q},\omega) d \omega =
  \frac{\pi} {4} (1-c_h). \label{eq8}
\end{equation}
The response should qualitatively correspond to a paramagnet close to
the AFM instability, so it 
is assumed of the form
\begin{equation}
  \chi''({\bf q},\omega) \propto\frac{ \phi(\omega,T)}{
(\tilde q^2 +\kappa^2) (\omega^2+\omega_\kappa^2)}, \label{eq9}
\end{equation}
where $\phi(\omega,T) \propto \omega$ would be appropriate for a
nearly AFM Fermi liquid \cite{pine,chub} or an undoped AFM in 2D.  The
selfconsistent set of equations for $G$ is closed with
$\Sigma=\Sigma_{\rm pm}+\Sigma_{\rm lf}$.  For given chemical
potential $\mu$ the FS emerges as a solution determined by the
relation $\zeta_{{\bf k}_F} + \Sigma'({{\bf k}_F},0)= \mu$.  We should
note that at given $\mu$, electron concentration $c_e$ as calculated
from the density of states ${\cal N}(\omega)=(1/N)\sum_{\bf k} A({\bf
k},\omega)$ integrated for $\omega<0$, does not in general coincide
with the one evaluated from the FS volume, $\tilde c_e=V_{\rm
FS}/V_0$.  Nevertheless, apart from the fact that within the $t$-$J$
model validity of the Luttinger theorem is anyhow under question
\cite{putt}, in the regimes of large FS both quantities appear to be
quite close. The position of the FS is mainly determined by
$\zeta_{\bf k}$ and $\Sigma_{\rm pm}$, while in this respect
$\Sigma_{\rm lf}$ is less crucial.
We choose further on parameters $J=0.3t, t'=-0.2t$ and $\kappa=
\sqrt{c_h}$ while 
$\eta_1$ and $\eta_2$ are determined as a function of $c_h$ from model
calculations \cite{bonca89}. We use $N=32\times32$ points in the
Brillouin zone and broadening $\epsilon/t=0.02$.

In Fig.~1 we present hole concentration $c_h$ vs the chemical
potential $\mu/t$ as obtained from Eq.~\ref{eq5}. We solve the
equation by iteration. With labels (a), (b) and (c) are indicated
special cases presented in next figures. At $c_h \sim 0.12$ we observe
in the equations an instability signaled by oscillatory behavior
instead of convergence and a unique solution can not be obtained in
the region indicated by the dashed line. However, at lower doping, $c_h <
0.05$, the solution converges again and a typical result is
indicated with (d). The region of instability coincides with the
transition from the large to a small FS, as presented below.

In Fig.~2 are presented spectral functions $A({\bf k},\omega)$ along
the principal directions in the Brillouin zone for cases (a), (c) and
(d) from Fig.~1.  It is evident that $\Sigma_{\rm pm}$ leads to strong
damping of hole QP and quite incoherent momentum-independent spectrum
$A({\bf k},\omega)$ for $\omega \ll -J$ which qualitatively reproduces
ARPES and numerical results \cite{jprev}.  Electron QP (at $\omega>0$)
are in general very different, i.e., with much weaker damping arising
only from $\Sigma_{\rm pm}$.  At low doping $c_h<0.05$ we find the
regime of small (pocket-like) FS.

The shape of the FS is most clearly presented with contour plot of the
electron momentum distribution function $\tilde n({\bf k})=
\alpha^{-1}\int_{-\infty}^0
A({\bf k},\omega){\rm d} \omega$.  
Results for a characteristic development of the
FS with $c_h$ are shown in Fig.~3. At $c_h <c_h^* \sim 0.05$ solutions
are consistent with a small pocket-like FS, (d), whereby this behavior
is enhanced by $t'<0$ as realized in other model studies \cite{tpr}.
On increasing doping the FS rather abruptly changes from a small into a
large one as suggested from the results of SCBA \cite{ram00}.  The
smallness of $c_h^*$ has the origin in quite weak dispersion dominated
by $J$ and $t'$ at $c_h
\to 0$ which is overshadowed by much larger $\zeta_{\bf k}$ at moderate
doping, where the FS is large and its shape is controlled by $t'/t$.
Figures (a), (b) and (c) correspond to higher doping with common large
FS topology. However, in the intermediate doping regime, (b) and (c),
the pseudogap is pronounced at momenta around $(\pi,0)$ point. The
gap is more pronounced in (c) because of longer AFM correlation length
$\xi$ (smaller $\kappa$).  In Fig.~4 the development of the PG is
presented for $c_h=0.21$ [case (b)]. Here $\kappa=0.4$
and the PG is not fully developed yet. With increasing $\xi$ -- for
this doping somewhat unrealistic values -- the gap opens. At lower
doping, case (c) in Fig.~3 the gap is opened what also shows up the
form of the FS which tends to avoid points $(\pi,0)$, $(0,\pi)$.

In conclusion, we have presented a theory for the spectral functions
within the $t$-$J$ model where the double-occupancy constraint is
taken explicitly into account and used to derive an effective
spin-fermion coupling. The coupling to transverse AFM paramagnons is
strong, nevertheless it can be well treated within a generalized
SCBA. On the other hand, the coupling to longitudinal AFM
fluctuations, $\tilde m_{\bf kq}$, is moderate near FS for low doping
and leads to a pseudogap. The latter is however not in contradiction
with the existence of a large FS, and should show up in integrated
photoemission and ARPES results as well as in the uniform susceptibility and
in the specific heat.

 \vskip 2 cm 
\centerline{F I G U R E S}

\noindent   
\begin{figure}[htb]      
\center{\epsfig{file=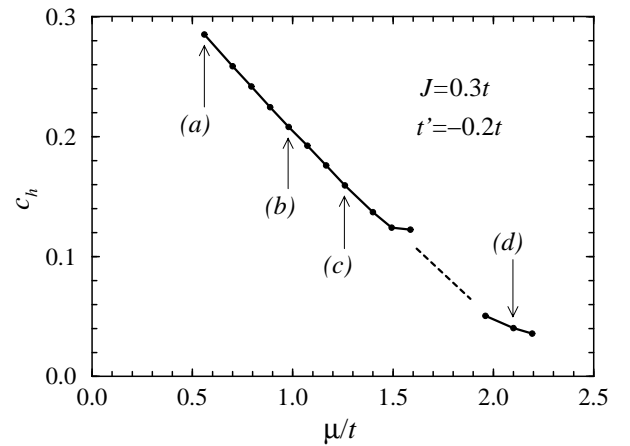,width=75mm,angle=-90,clip=}}
\caption{Hole concentration $c_h$ as a function of the chemical potential
$\mu/t$ for $J/t=0.3$, $t'/t=-0.2$ and $N=32 \times 32$ sites in the
Brillouin zone. Note selected cases presented 
further on: (a) $c_h=0.28$, (b) $c_h=0.21$, (c) $c_h=0.16$ 
and (d) $c_h=0.04$.} 
\end{figure}

\newpage\clearpage
\widetext
\noindent   
\begin{figure}[htb]   
\center{\epsfig{file=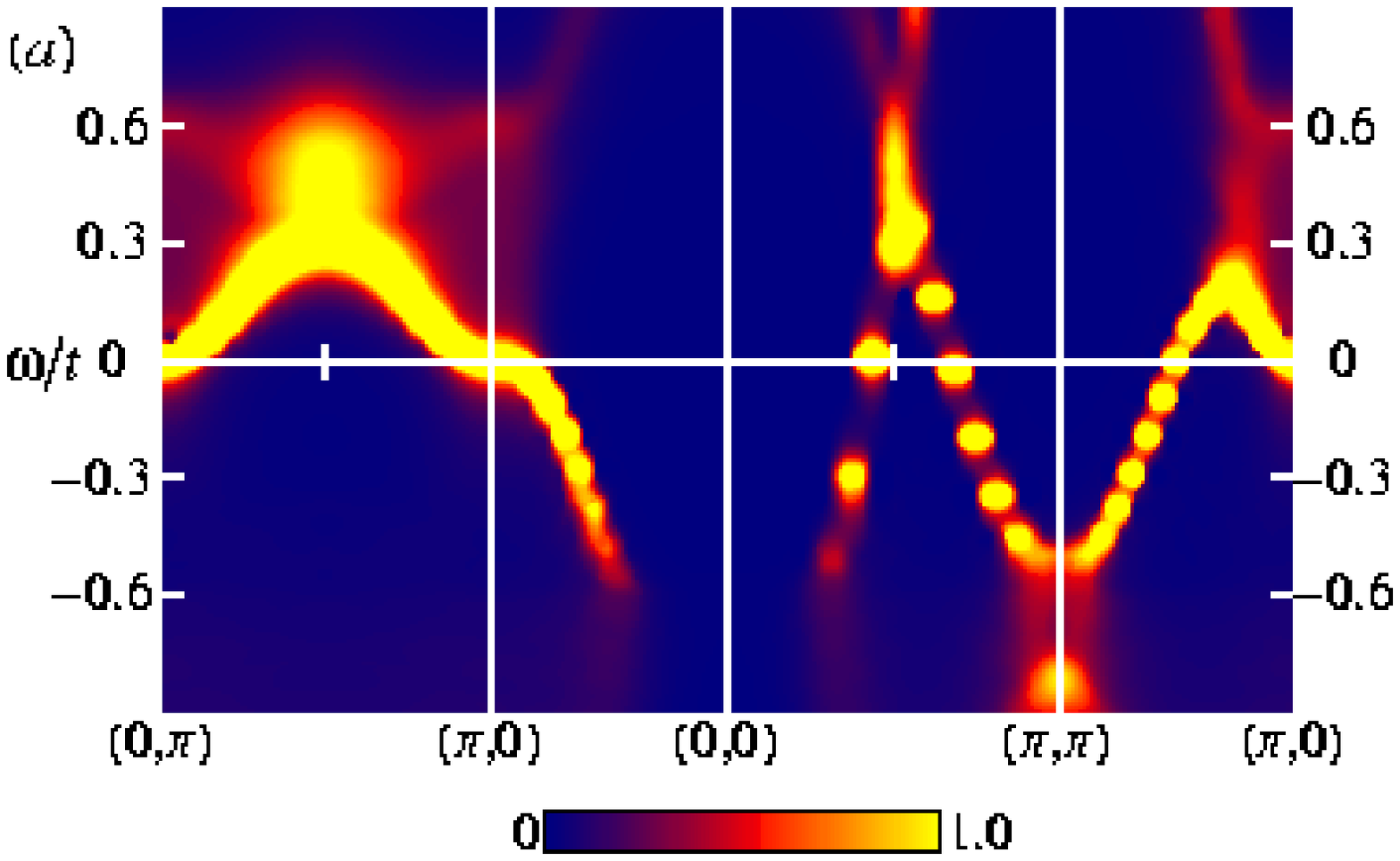,width=90mm,angle=-0,clip=}}\\[-11mm]
\vskip -11 mm
\center{\epsfig{file=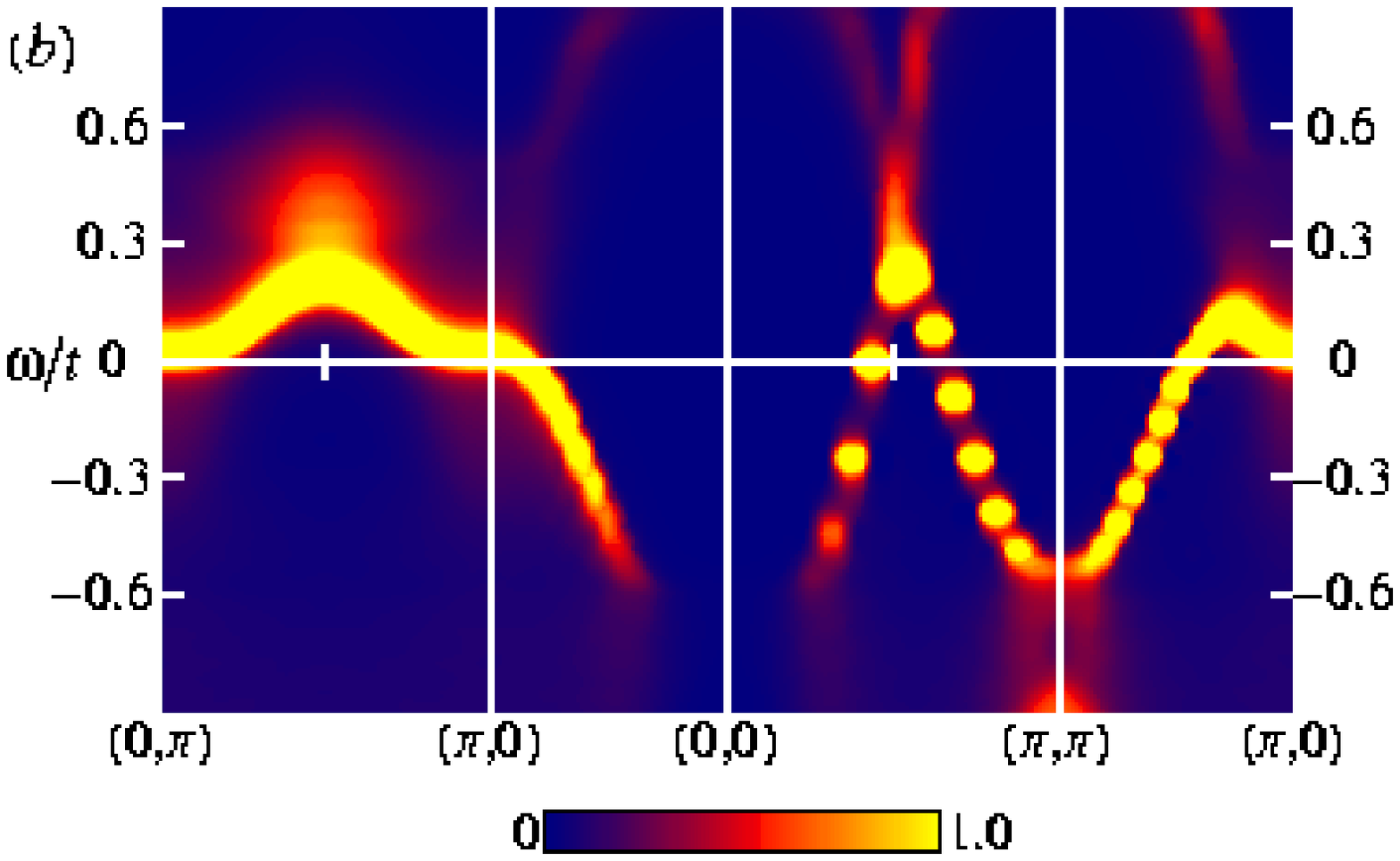,width=90mm,angle=-0,clip=}}\\[-11mm]
\vskip -11 mm
\center{\epsfig{file=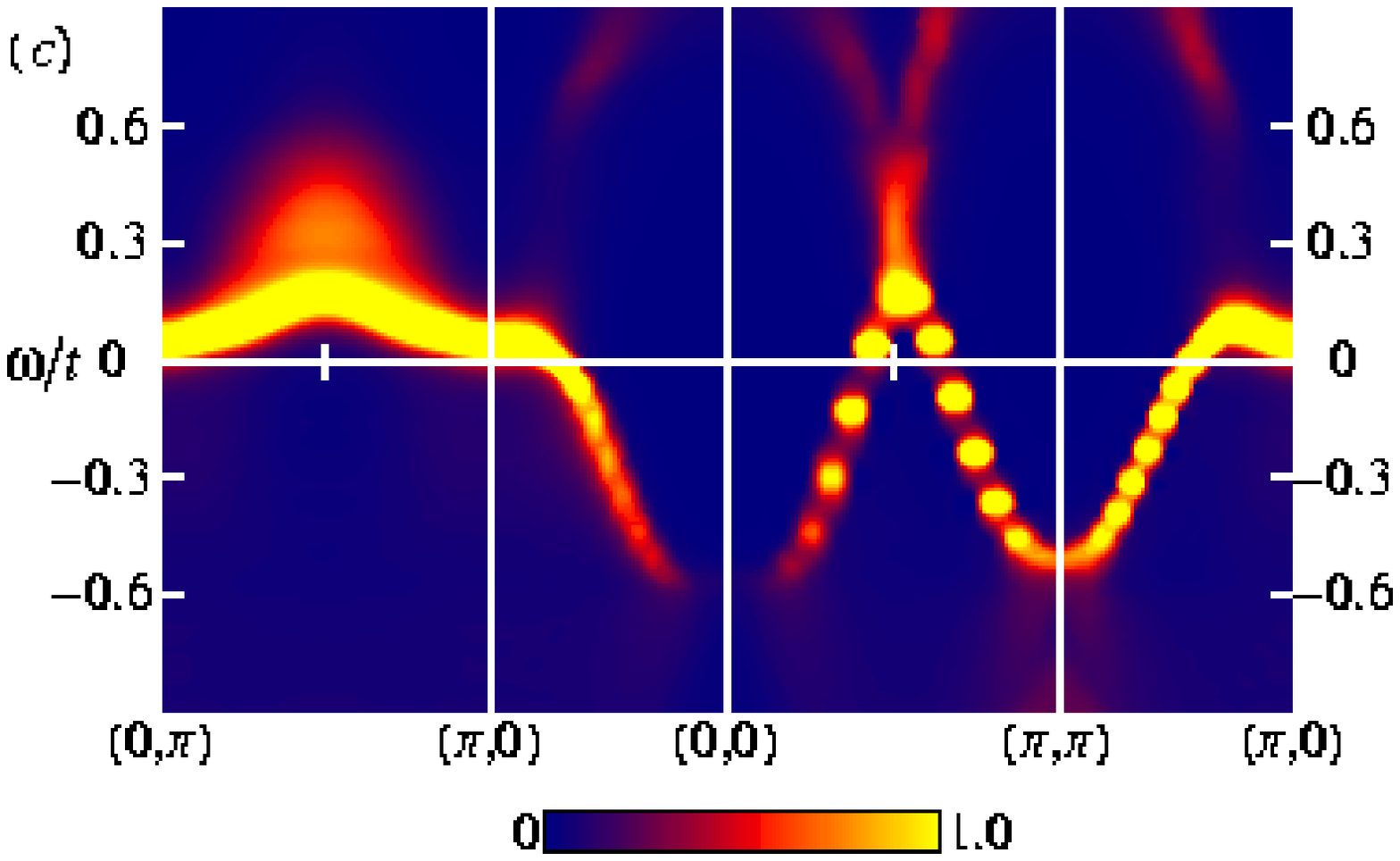,width=90mm,angle=-0,clip=}}\\[-11mm]
\vskip -11 mm
\center{\epsfig{file=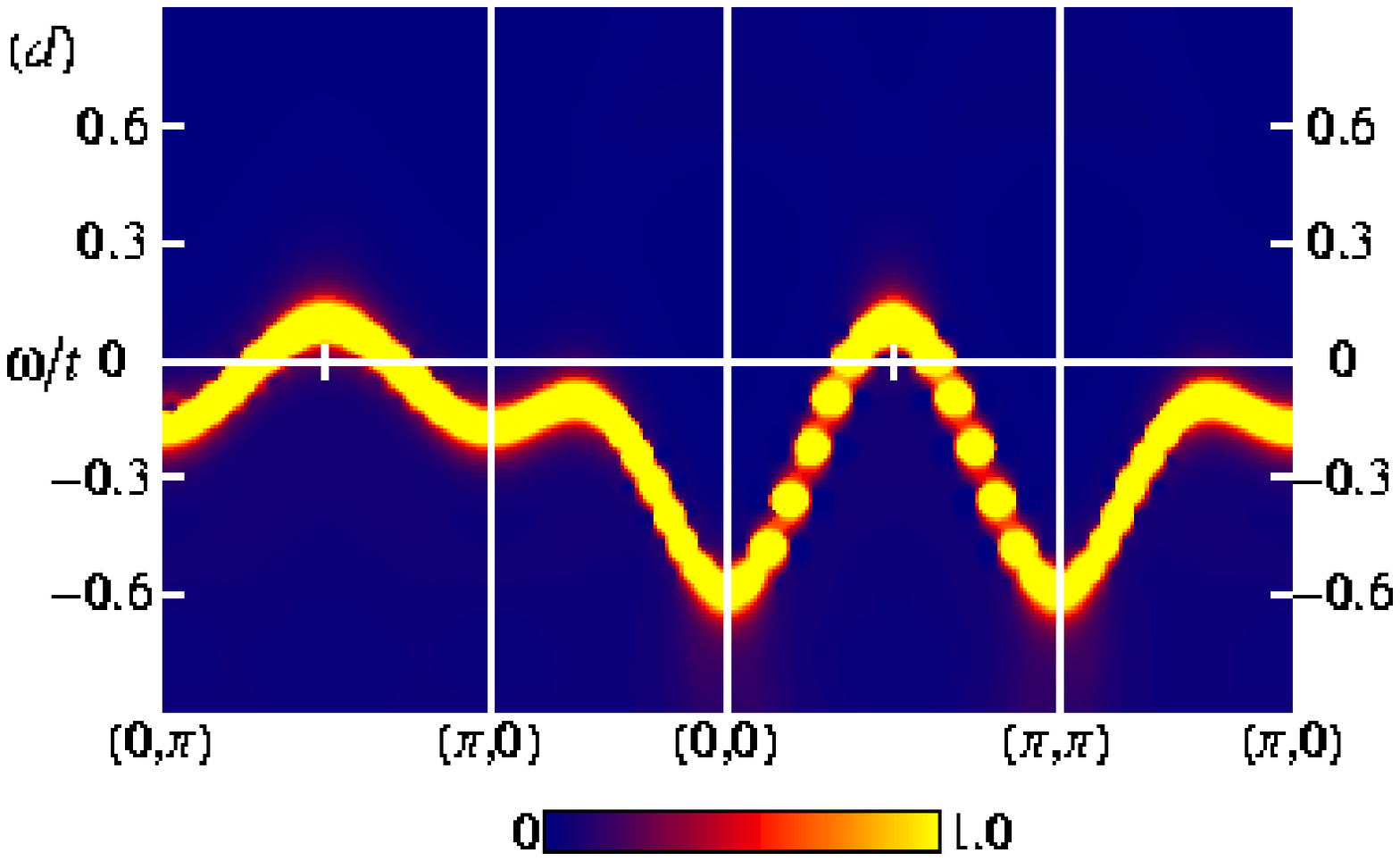,width=90mm,angle=-0,clip=}}  
\caption{(color) Spectral functions $A({\bf k},\omega)$ along principal lines
in the Brillouin zone. For convenience of presentation the functions
are clipped at $A({\bf k},\omega)=1.0$. Labels (a) -- (d)
correspond to selected cases in Fig.~1.} 
\end{figure}

\newpage\clearpage
\noindent       
\begin{figure}[htb]   
\center{\epsfig{file=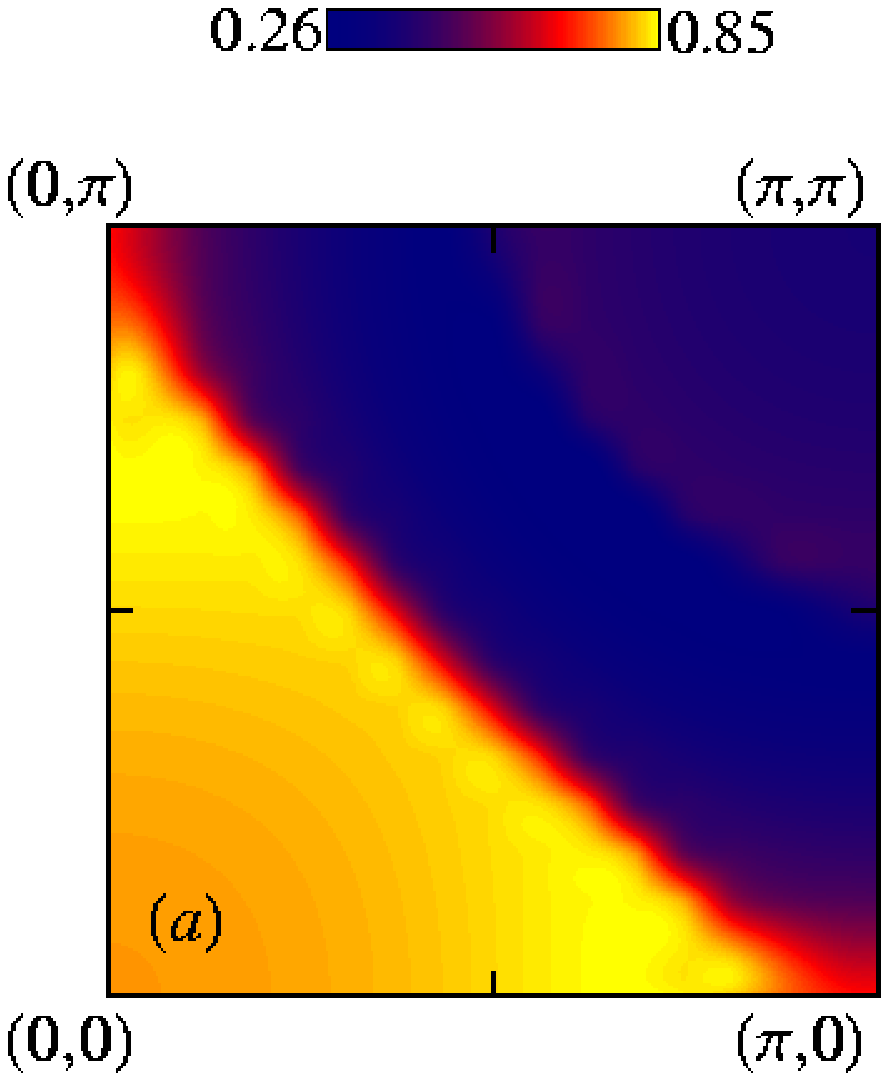,width=70mm,angle=-0,clip=}
\epsfig{file=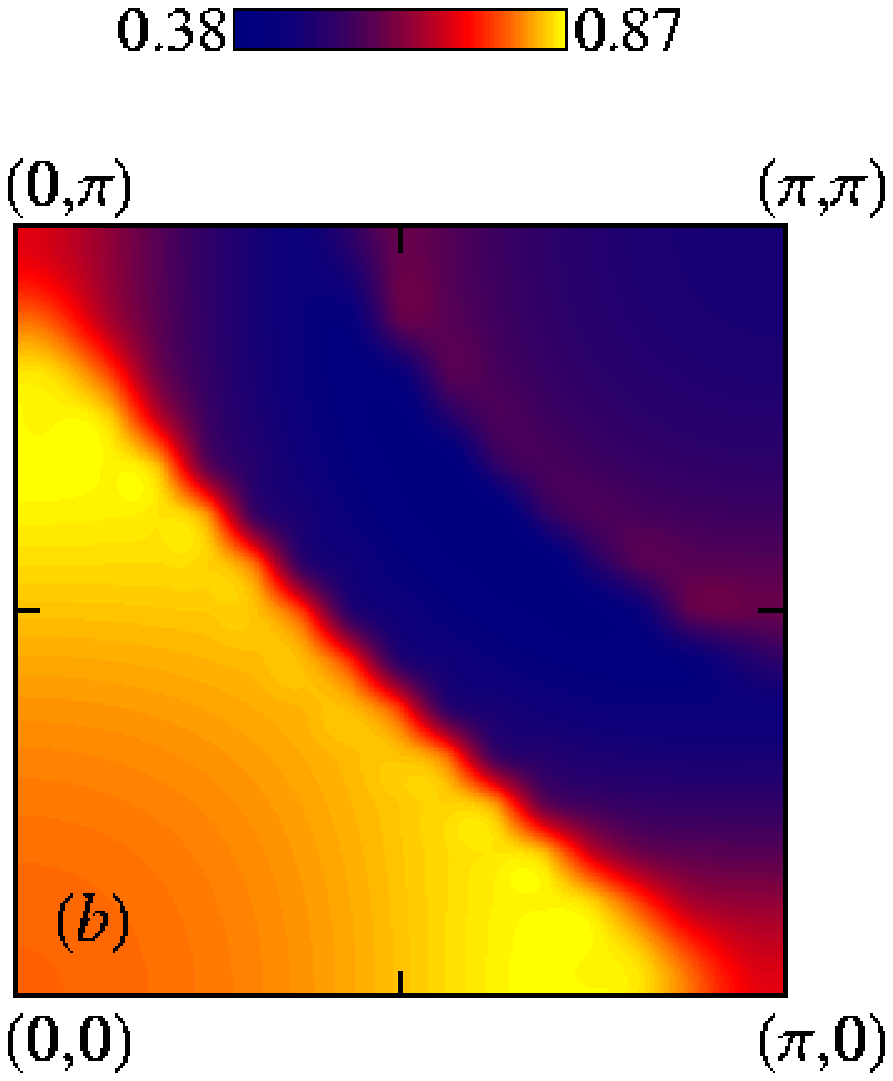,width=70mm,angle=-0,clip=}}
\center{\epsfig{file=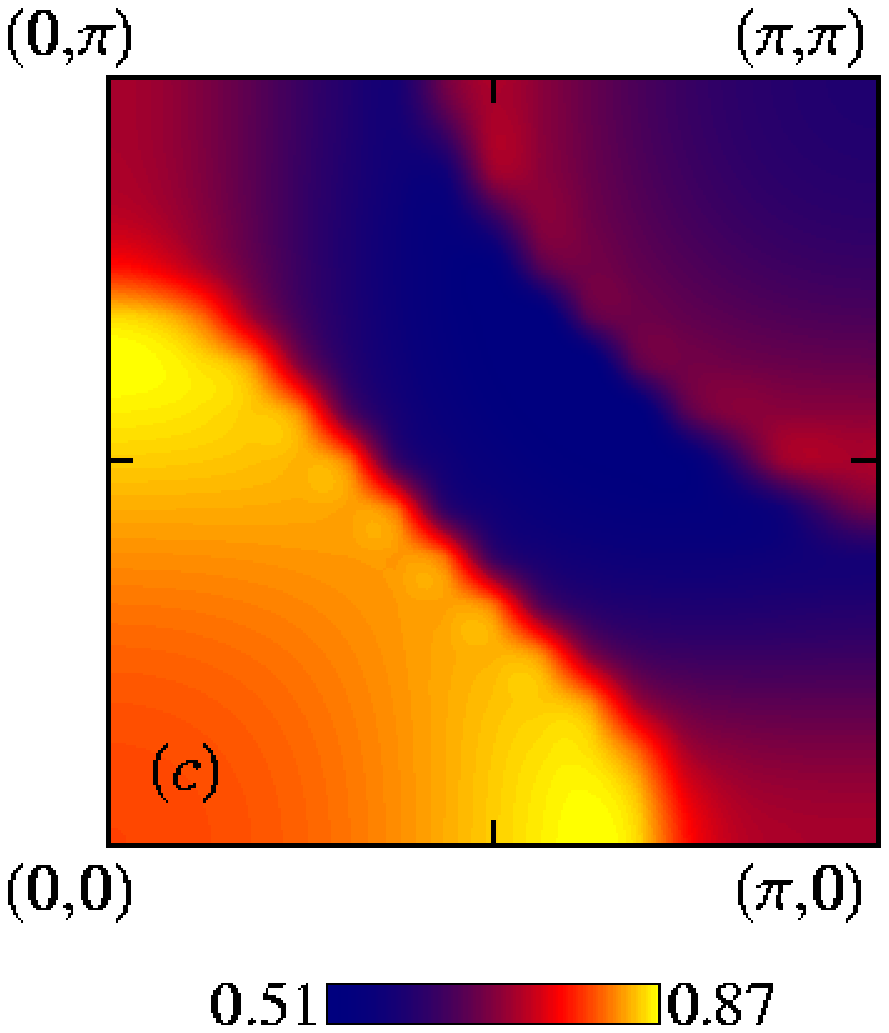,width=70mm,angle=-0,clip=}
\epsfig{file=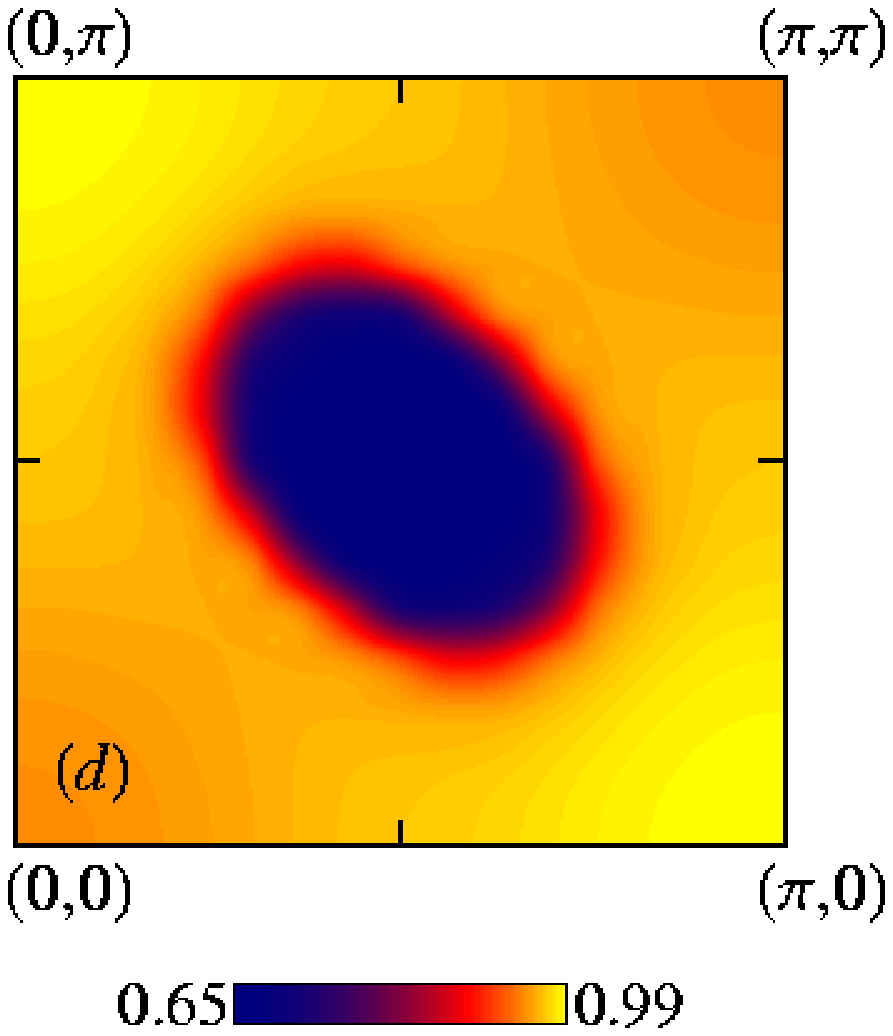,width=70mm,angle=-0,clip=}}
\caption{(a,b,c,d) (color) Electron momentum 
distribution function $\tilde n({\bf k})$ for various
$c_h$ as selected in Fig.~1.} 
\end{figure}

\newpage\clearpage
\noindent   
\begin{figure}[htb]   
\center{\epsfig{file=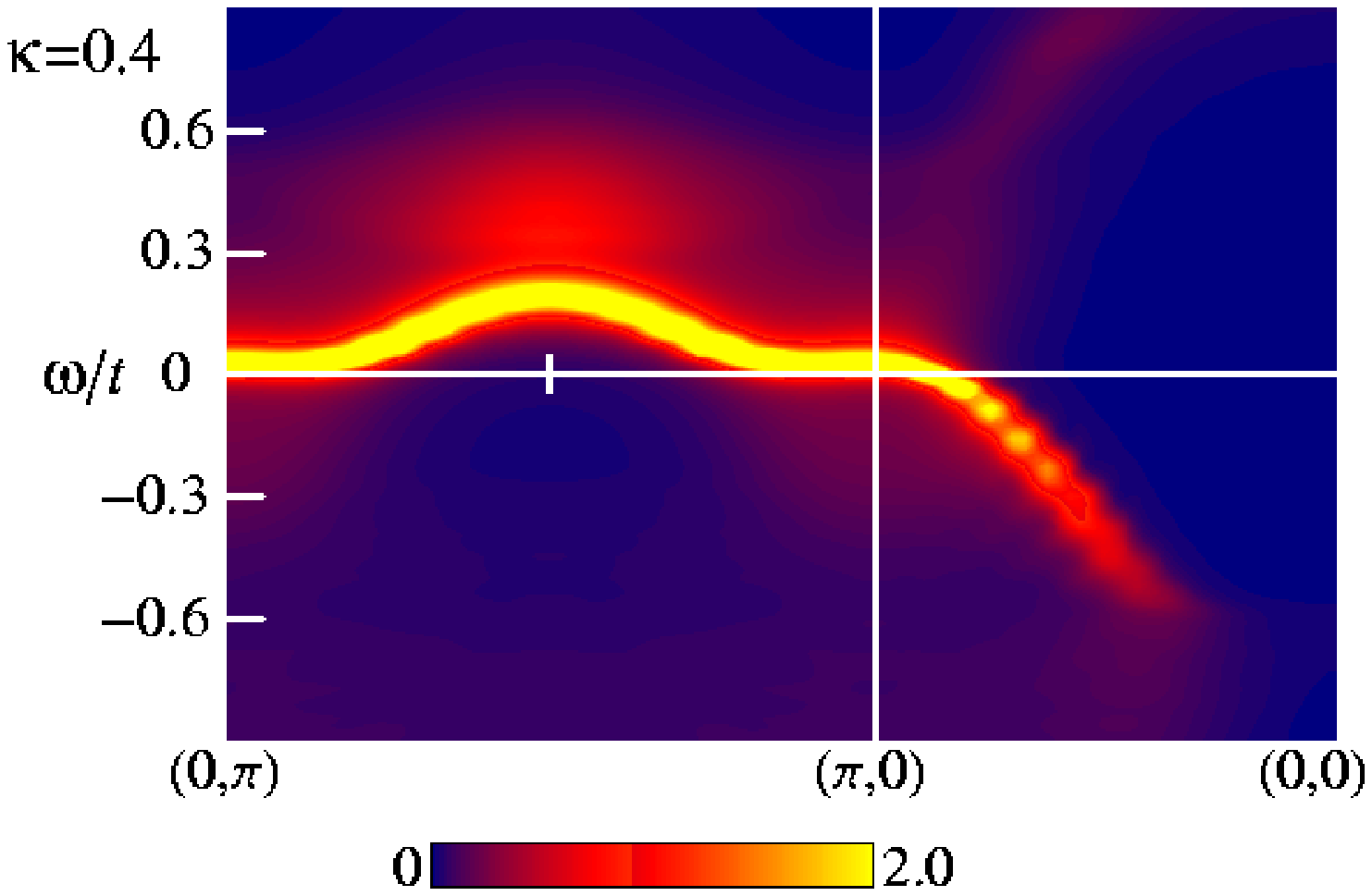,width=110mm,angle=-0,clip=}}\\[-12mm]
\vskip -12 mm
\center{\epsfig{file=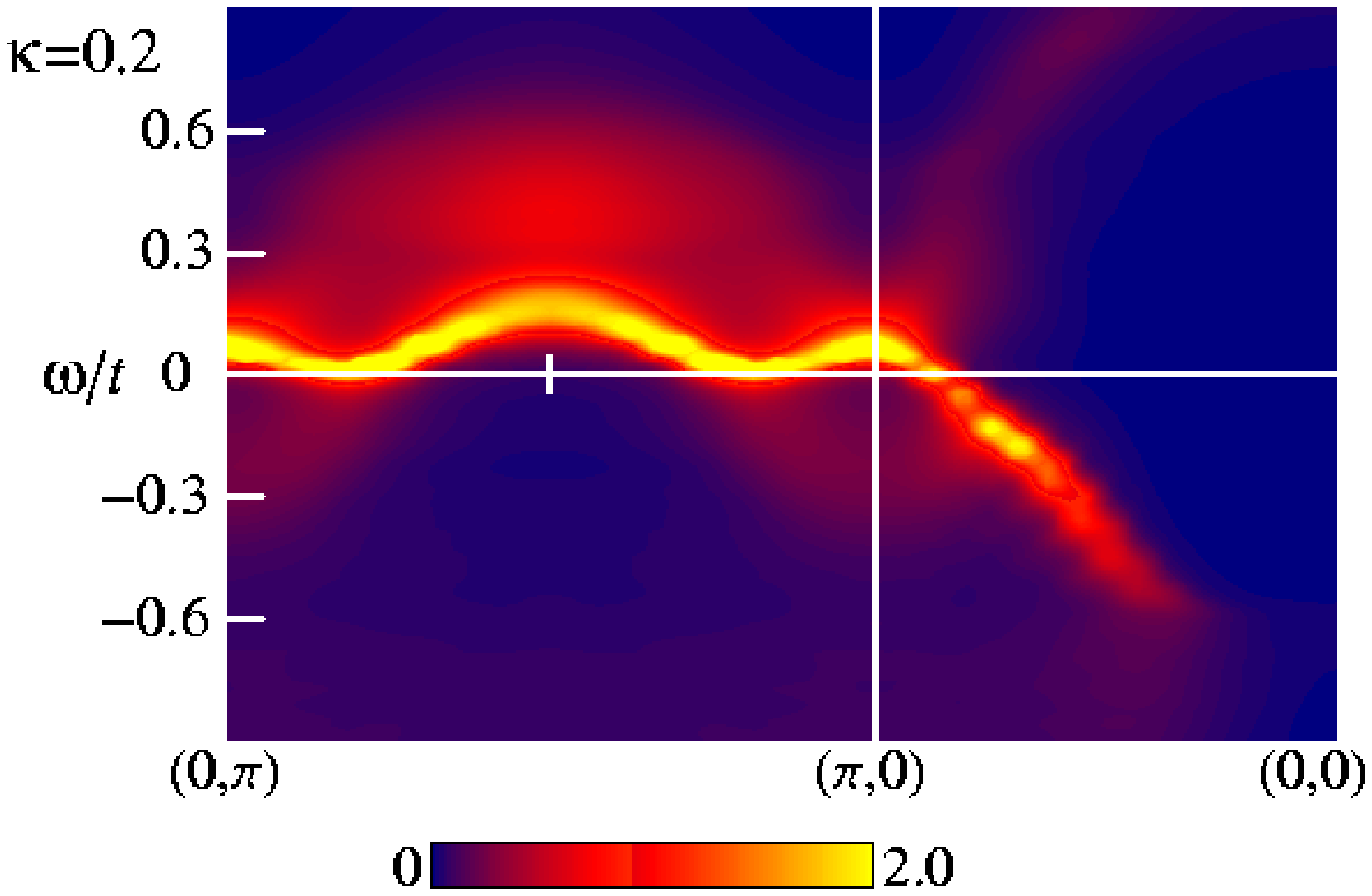,width=110mm,angle=-0,clip=}}\\[-12mm]
\vskip -12 mm
\center{\epsfig{file=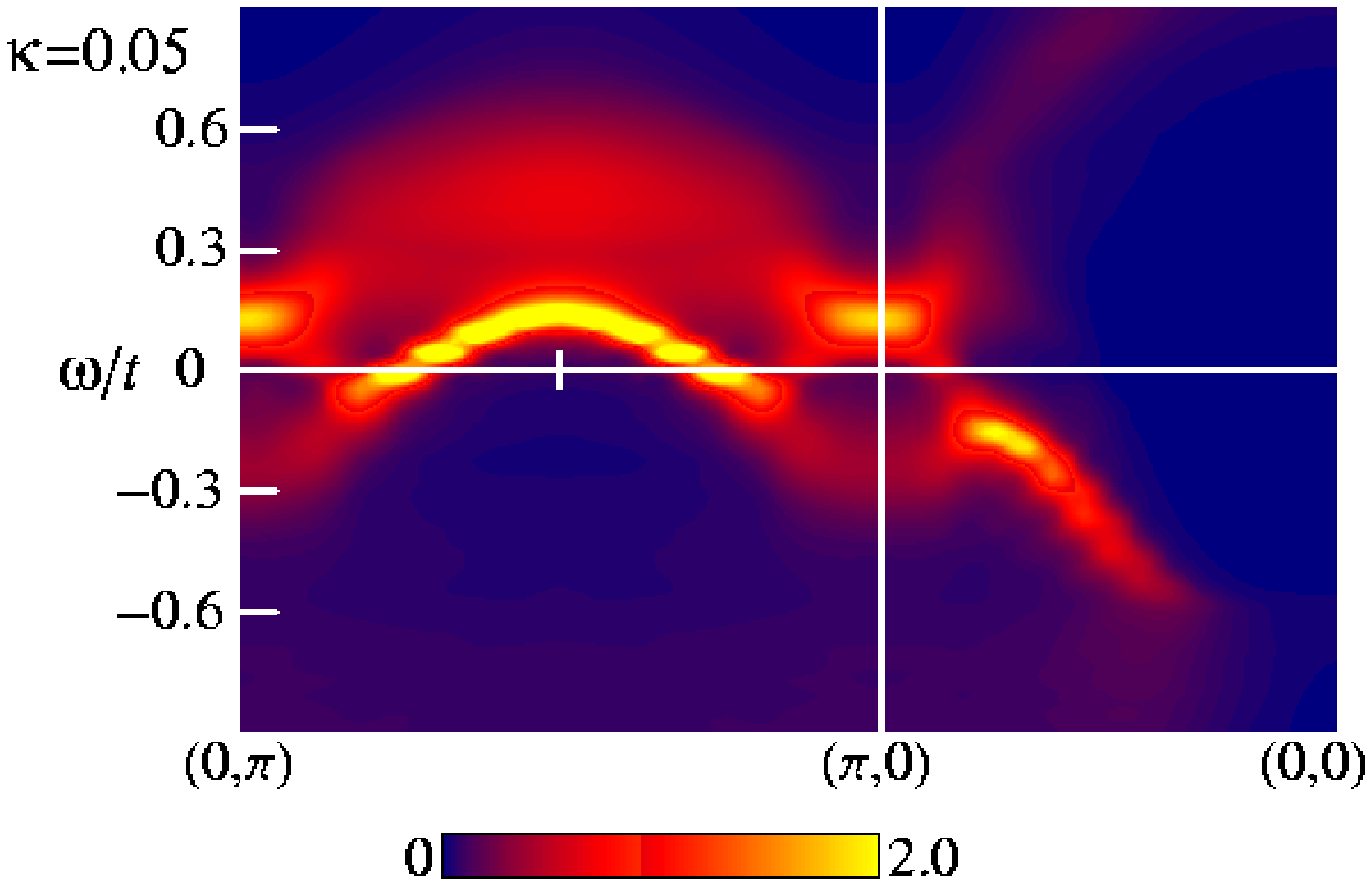,width=110mm,angle=-0,clip=}}
\caption{(color) Spectral functions $A({\bf k},\omega)$ for $c_h=0.21$ [(b) in
Fig.~1] for different $\kappa$. 
The functions are clipped at $A({\bf k},\omega)=2.0$.
Note the opening of the gap at ${\bf
k}=(\pi,0)$ with decreasing $\kappa$.} 
\end{figure}

\end{document}